# Rapid prototyping and performance evaluation of MEC-based applications

Alessandro Noferi[1], Giovanni Nardini[2], Giovanni Stea[2], Antonio Virdis[2]
Dipartimento di Ingegneria dell'Informazione, University of Pisa
Largo L.Lazzarino, 1, 56122, Pisa, Italy
[1]alessandro.noferi@phd.unipi.it, [2]{name.surname}@unipi.it

*Abstract*— Multi-access Edge Computing (MEC) will enable context-aware services for users of mobile 4G/5G networks. MEC application developers need tools to aid the design and the performance evaluation of their apps. During the early stages of deployment, they should be able to evaluate the performance impact of design choices - e.g., what round-trip delay can be expected due to the interplay of computation, communication and service consumption. When a prototype of the app exists, it needs to be tested it live, under controllable conditions, to measure key performance indicators. In this paper, we present an open-source framework that allows developers to do all the above. Our framework is based on Simu5G, the OMNeT++-based simulator of 5G (NewRadio) and 4G (LTE) mobile networks. It includes models of MEC entities (i.e., MEC orchestrator, MEC host, etc.) and provides a *standard-compliant* RESTful interface towards application endpoints. Moreover, it can interface with external applications, and can also run *in real time*. Therefore, one can use it as a *cradle* to run a MEC app live, having underneath both 4G/5G data packet transport and MEC services based on information generated by the underlying emulated radio access network. We describe our framework and present a use-case of an emulated MEC-enabled 5G scenario.

## 1. Introduction

Multi-access Edge Computing (MEC) will deliver cloud-computing capabilities at the edge of the network. Besides providing a smaller and more predictable latency, this will enable context-awareness, capitalizing network information such as radio access conditions, user location, etc., which would be difficult to know, if not impossible, in a cloud-based application. This is achieved by having MEC communicate with the access network, via the so-called *MEC services*. These are services exported by a MEC platform connected to the access technology, accessible via a RESTful interface, that a MEC application (MEC app, henceforth) can query to acquire the user context. They provide information on the user (e.g., its own radio conditions or location) as well as on the access itself (e.g., what is the current network load or number of users), thus allowing one to create advanced user services. Examples of these user services – already discussed in the context of the European Telecommunications Standards Institute (ETSI), who is standardizing MEC – are: user QoS prediction based on current radio conditions, location and movement pattern; vehicular alerts, e.g. a car approaching a slippery patch of tarmac or robot swarm coordination in a factory.

While, as its very name suggests, MEC is access-technology agnostic, it is quite clear that its interplay with mobile networks, namely 4G and 5G cellular ones today – and 6G tomorrow – will be prominent in the future. This is because ubiquitous, regulated, reliable and secure wireless access is a pre-requisite for marketable user services. In this scenario, MEC services are provided by the underlying 4G/5G radio access, leveraging the information base available at the base stations (eNB in LTE, gNB in New Radio) and in the various network entities involved in the control and management plane. In this paper we refer to this scenario, without explicitly repeating it henceforth.

By making the hosting infrastructure open – under controlled conditions – MEC enables the role of *MEC app developer*. This will not necessarily be related to either the network operator or the MEC infrastructure provider. MEC apps developed by this player will be hosted in the MEC infrastructure, and they will communicate with the user (on behalf of whom they are run) through the cellular network run by the operator. More to the point, the MEC app will also interact with the cellular network by consuming MEC services provided by it. It is foreseen that the growth of the community of MEC app developers will be the driving force for a widespread MEC diffusion.

In this scenario, it becomes particularly important for MEC app developers to be able to test their applications for functionality and performance in credible and controllable settings. By *credible*, we mean taking into account: i) radio resource contention on the data plane of the underlying cellular network; ii) resource contention at the MEC host, given that MEC apps run as virtual machines or containers on a shared computing infrastructure, and iii) contention for access to MEC services (e.g., queueing delays), due to concurrent service requests. These settings must be fully controllable, e.g. allow one to test the impact of radio congestion, poor channel, MEC platform congestion, etc., especially if the user services are time-critical (as are some of those mentioned at the beginning of this section) or come with service level agreement guarantees. Functional and performance testing of MEC apps is required at different stages in the development cycle: early on, one may want to understand the performance implications of alternative designs, e.g., the ensuing communication patterns, before committing to a given solution. Later, one may want to test a prototype of its MEC app in real time, measuring delays and possibly factoring in user interaction. Eventually, a developer may want to showcase its own MEC app, completed and polished, in a live demo, e.g. for sales pitch to prospective financers. All this is made difficult by the lack of suitable tools, and the fact that it is hardly possible to use a live 5G network. There are indeed tools that simulate end-to-end communications on cellular networks – see, e.g., [16][17], tools that simulate the MEC environment – see, e.g., [23] and survey [24], open-source environments to host MEC apps – see, e.g. [28]. There are also tools that allow one to test MEC service interfaces – see, e.g., [41][42]. However, none of these can work in synergy, hence – while certainly helping a developer – they do not allow the kind of support to prototyping discussed above.

In this paper we describe an open-source complete framework to integrate all the above functionalities. It is based on Simu5G [6][7], an open-source simulation library for 5G (and 4G) cellular networks based on OMNeT++ [8]. Our framework models the whole MEC infrastructure – including the MEC orchestrator, the MEC host running MEC apps and the MEC platform providing MEC services – in an environment devised for end-to-end, application-oriented discrete-event simulation. All the interfaces between the MEC environment and the application world are designed to be ETSI-compliant. Moreover, our framework can interface with external application endpoints and carry their traffic, and can do so *in real time* [2][3], acting as a network emulator [4][5]. A MEC app developer can thus exploit it as a *cradle* for distributed MEC-based applications, which can be both tested for functional compliance and evaluated for performance, maintaining full control over the experimental setup. For instance, a developer may want to test a UE app (the one residing on the user device) with a stub of a MEC app (residing on the MEC host), which she can quickly develop as an OMNeT++ module within Simu5G. Alternatively, she may want to test a MEC app that consumes MEC services (e.g. the Radio Network Interface Service – RNIS [31], or the Location Service [32]), or find under what load conditions the response time of said MEC services may affect the end-to-end application performance. In this case, the MEC app could be developed within Simu5G, or outside it, e.g., hosted on a real MEC host such as Intel OpenNESS [28], consuming MEC services provided by Simu5G. For instance, developers of an edge-assisted Virtual Reality application [10] may be interested in assessing the Quality of Experience perceived by users when the application runs in a MEC system and exploits context information from a realistic 5G network scenario.

Two comments are in order. The first one is that – as far as we know – there seems to be no tools comparable to the one described in this paper. If you want to perform a live run of your MEC app in a 5G network, it seems that the only alternative is that you avail yourself of a 5G network testbed *and* a MEC infrastructure connected to it, with MEC services enabled, and run your app over them. This is expensive, time consuming and might still not allow you full control over experimental conditions (say, to create congestion or radio impairments). Our second comment is that the work presented in this paper can be expected to be useful also to the *other* players involved in the MEC chain of value as well. On one hand, a cellular network operator will need to test the impact of MEC-generated traffic on its Radio Access Network data plane, as well as the impact of MEC service requests on its network's control/management plane. On the other hand, MEC infrastructure providers may want to use these tools to plan for capacity, evaluate the performance impact of a growing number of hosted MEC apps on the MEC platform, identify bottlenecks and enforce admission control, etc.. Moreover, this work is currently being used to support the demonstration of federated learning of eXplainable Artificial Intelligence (XAI) models within 6G flagship EU project Hexa-X [29].

We describe the modeling of the MEC subsystem within Simu5G and show how these models are built for scalability, i.e., allow one to simulate congestion at the MEC platform efficiently, so that they still allow one to emulate in real time a large-scale computation and communication scenario (hundreds of simultaneous MEC apps, complex 5G network with tens of nodes and hundreds of users) on a desktop PC. Moreover, we describe an example of a MEC-based application running live through our framework, showing that setting up a testbed is quite simple and requires inexpensive hardware resources.

The rest of this paper is organized as follows: Section 2 provides background information on the technologies we use. Section 3 and Section 4 present the architecture of our software framework, focusing on the model of the MEC infrastructure and the MEC applications and services, respectively. Section 5 presents a use case. Section 6 reviews the related work. Finally, Section 7 draws conclusions and highlights directions for future work.

## 2. Background

In this section, we provide background knowledge required to understand the design choices for our MEC prototyping framework. More specifically, we present the capabilities of the Simu5G OMNeT++-based library, and we introduce the reference architecture of ETSI MEC.

### 2.1. Overview of Simu5G

Simu5G [6][7] is the evolution of the well-known SimuLTE 4G network simulator [25] towards 5G NewRadio (NR) access. It simulates the data plane of both the core and the radio access networks. Since it incorporates all SimuLTE's functionalities, it allows users to create legacy or mixed 4G/5G scenarios as well. Hereafter, we describe those entities and functionalities of Simu5G that are more closely related to the scope of this paper. We refer the interested reader to [6][7] for more details.

Simu5G is a *model library* for the OMNeT++ discrete-event simulation framework [8]. The latter allows analyzing any kind of system in which there are entities communicating with each other. In OMNeT++ these entities are represented by *simple modules* communicating via message exchange through connections among module gates. Module's behavior is coded in C++. Multiple modules can be composed to form a *compound* module. Henceforth, we will refer to modules to denote either simple or compound OMNeT++ modules. Simu5G is interoperable with all the libraries based on OMNeT++, such as INET [9] for TCP/IP-based network technologies, and Veins [11] for vehicular mobility. This allows users to construct very complex scenarios straightforwardly, by importing and connecting existing libraries, with hardly any extra line of code required. Simu5G, in fact, leverages several models and functionalities taken from INET itself. Figure 1 shows the main modules of Simu5G.

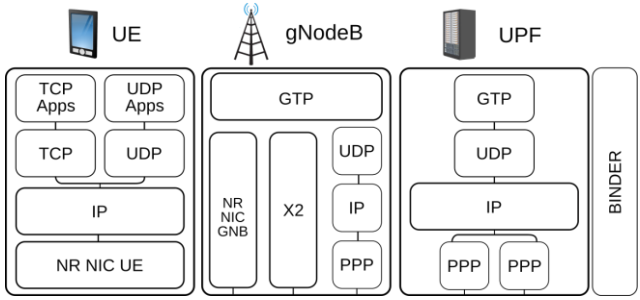 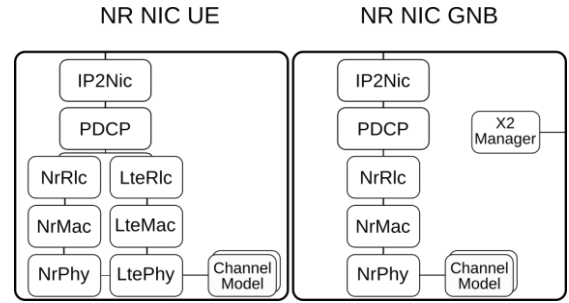

Figure 1 - Main modules of the Simu5G model library

Figure 2 - Internal model of the NR NIC

As far as the core network (CN) is concerned, Simu5G defines a User Plane Function (UPF) or Packet GateWay (PGW) and allows users to construct arbitrary CN topologies, where packet forwarding is based on the GPRS tunneling protocol (GTP). For what regards radio access, Simu5G defines gNBs and UEs, which communicate using the New Radio protocol stack at layer 2. gNBs can be connected to the CN directly, in the so-called StandAlone deployment, or operate in an E-UTRA/NR Dual Connectivity (ENDC) deployment, where LTE and NR coexist. In this last configuration, the gNB works as a secondary node for an LTE eNB, which acts as master node connected to the CN [12].

Both UEs and gNBs include a NR Network Interface Card (NIC), which models the NR protocol stack. With reference to Figure 2, the NR NIC includes all the sublayers of NR (from the Packet Data Convergence Protocol to the Physical Layer), with 3GPP-compliant behavior. As for the physical layer, Simu5G follows the approach already used by SimuLTE, i.e. to model the *effects* of propagation on the wireless channel at the receiver, *without* modelling symbol transmission and constellations. This allows us to compute the Signal-to-Interference-and-Noise Ratio (SINR) at receivers correctly and to compute MAC-layer decoding probabilities accordingly, at a manageable complexity, and makes the simulator more evolvable.

Simu5G simulates radio access on multiple carriers, in both Frequency- and Time-division duplexing (FDD, TDD). Different carrier components can be configured with different FDD numerologies and different TDD slot formats. Moreover, different carrier components can have different channel models and MAC-level schedulers. Simu5G incorporates UE handover and network-controlled device-to-device communications, both one-to-one and one-to-many.

Simu5G can run in real time, thanks to OMNeT++ and INET functionalities. In fact, OMNeT++ allows events to be scheduled by a *real-time scheduler*, which matches the simulated time to the wall-clock time at the beginning of a simulation, and slows down event scheduling until the corresponding wall-clock time has expired. This is only possible, of course, if simulated time would otherwise run *faster* than wall-clock time, something which depends on both the hardware/software system that runs OMNeT++, *and* the scenario being simulated. More specifically, the larger the *scale* of the simulation (e.g., the higher the number of UEs and gNBs), the less likely it is that this is possible. Parallel to this, INET allows one to endow modules with *external interfaces* that exchange packets with the outside world. This means that an

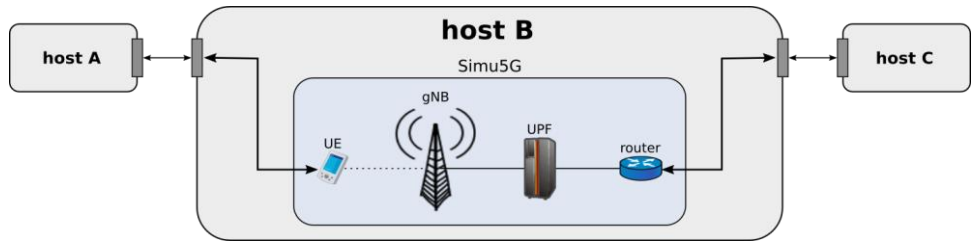
Figure 3 - Simu5G as emulator

external application can exchange packets with a module in a simulated network. Since simulation can also run in real time, this allows one to perform *real-time network emulation*, transporting packets between external application endpoints.

Figure 3 shows a simple scenario wherein host A and C are physically connected to host B, which runs an emulated scenario in Simu5G. Packets sent by host A appear in the emulation at the UE, and packets leaving the rightmost interface of the router are sent to host C. Packets flows similarly in the other direction as well. Both the UE and the router must be endowed with external interfaces for this to happen. Provided that routes are added to host B's operating system, an application on host A can thus reach any entity having an IP address in the emulated scenario (e.g., the UPF). Moreover, it can reach host C, with its packets traversing the emulated scenario. More complex configurations can also be created, such as having one remote host reachable using a public IP address.

Simu5G inherits the above capabilities – being based on OMNeT++ and INET – but is also explicitly designed for scalability, i.e., to allow real-time emulation of relatively large scenarios on a desktop machine. As shown in [3], it offers multiple models of entities (i.e., UEs and gNBs), which can be used for different purposes: *foreground* entities run the entire protocol stack and can be connected to external interfaces. On the other hand, *background* entities run only those functions that are necessary to create communication impairments: more specifically, background UEs and gNBs create lifelike interference at the physical layer, and background UEs contend for access to resources at the MAC layer. Background entities, however, have a reduced protocol stack and do not transmit packets over the air, hence cannot host external interfaces. Background entities create the same impairment that complete entities would, at a much lower overhead, which allows one to extend the scale of scenarios that can be emulated in real time. Work [3] shows that one can emulate scenarios with tens of gNBs and several hundreds of UEs, where two endpoints exchange several Mbps' worth of traffic, all on a desktop PC.

### 2.2. ETSI MEC Architecture

MEC is a computing infrastructure that can run applications (MEC apps) as virtual machines or containers on behalf of network users, interfacing with the access network (e.g., a cellular network). The MEC architecture has been standardized by the Industry Specification Group of ETSI [34]. The above standard covers the functional entities of the MEC architecture, as well as the procedures for a user to instantiate a MEC app. The data-plane exchange between a user and a MEC app is not part of the standard. This section describes the entities that compose the MEC architecture, with a focus on the ones implemented in the framework, which are highlighted in black in Figure 4. The MEC architecture includes a *MEC host* level and a *MEC system* level. The former contains the virtualization infrastructure used to run the virtual machines containing MEC apps. Within it, the MEC platform maintains a Service Registry, i.e. a catalogue that specified which *MEC services* can be used by MEC apps, and where they are located (i.e., at the MEC host itself, or at a different one). Examples of standardized MEC services are the Radio Network Information Service (RNIS) and the Location Service. Every MEC service advertises its presence to all the Service Registries in a MEC system via the MEC platform manager.

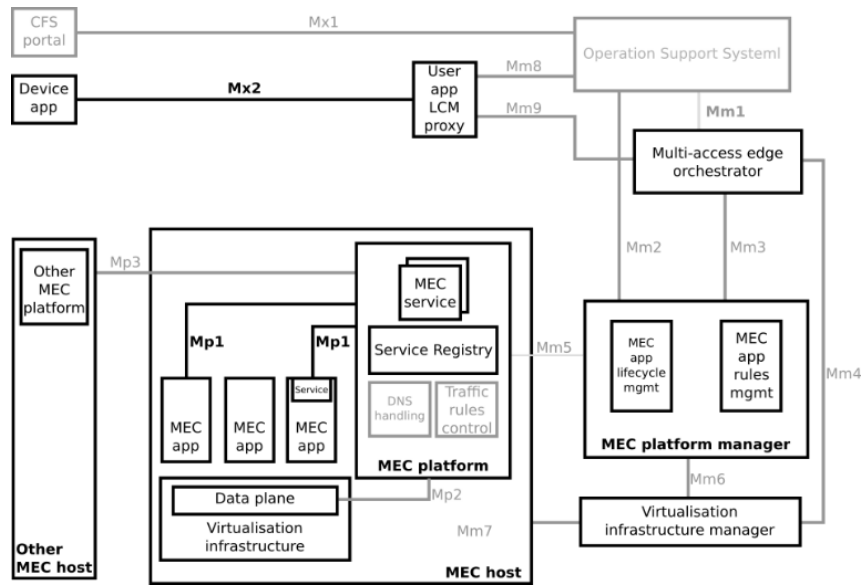

Figure 4 - Main entities in an ETSI-MEC infrastructure (taken from [34]). Black/grey modules are/are not modelled in our framework

When a MEC app queries a Service Registry, the latter returns the (local or remote) endpoints at which the available MEC service should be contacted. This allows MEC apps to consume MEC services located in different MEC hosts.

MEC apps can use MEC services via a standard MEC APIs, built upon RESTful APIs. REST follows a request-response paradigm. While in the MEC environment a MEC app often acts as a client, making requests, it is sometimes useful to allow the MEC service to notify MEC apps of some events at specific times, in a subscribe/notify approach, still implemented based on the RESTful pattern [33]. With this approach a MEC app registers its interest in a particular resource (subscription phase), and then, when an event occurs concerning that resource, the MEC service sends a notification to all the subscribers (notification phase).

The MEC system level maintains a global view of the MEC hosts in a MEC system, arbitrates MEC-host resources, and manages the lifecycle of the MEC apps, i.e. instantiation, relocation and termination. Its core element is the MEC orchestrator. The orchestrator receives the requests for instantiation or termination of MEC apps issued by the user's *device application*, after a granting operation made by the Operation Support System of the network operator, which is usually managed by the network operator and deals with authentications and authorizations. It then selects an appropriate MEC host for that MEC app, based on required constraints (e.g., latency), available resources (e.g., memory, disk, CPU), and available MEC services. Then it instructs the appropriate MEC platform manager to deploy the virtual machine that will run the MEC app. The User application lifecycle management proxy (UALCMP) acts as a bridge between the device application and a MEC system: the requests coming from the former are forwarded to a MEC orchestrator, and responses from the MEC orchestrator are sent to the device application. A MEC deployment can include several *MEC systems*, i.e., orchestrators managing disjoint sets of MEC hosts.

A MEC app is onboarded through an application package. The latter is composed of a bundle of files provided by the application provider, onboarded into the MEC system and used by the latter to instantiate an application. It also includes the Application Descriptor describing the application requirements and rules required by the MEC app [35].

A UE interacts with the MEC system via two logically distinct entities: the device app and the UE app. The device app interfaces with the MEC system to request specific functions related to life-cycle management of a MEC app (notably, instantiation or termination). The UE app is instead the endpoint of data-plane communication between the UE and the MEC app, which starts after the device app receives confirmation of the instantiation of the MEC app. The sequence diagram of a MEC app instantiation is shown in Figure 5. We remark that neither the interface between the UE app and the device app, nor the data-plane communication between the UE app and the MEC app, are part of the ETSI MEC standard.

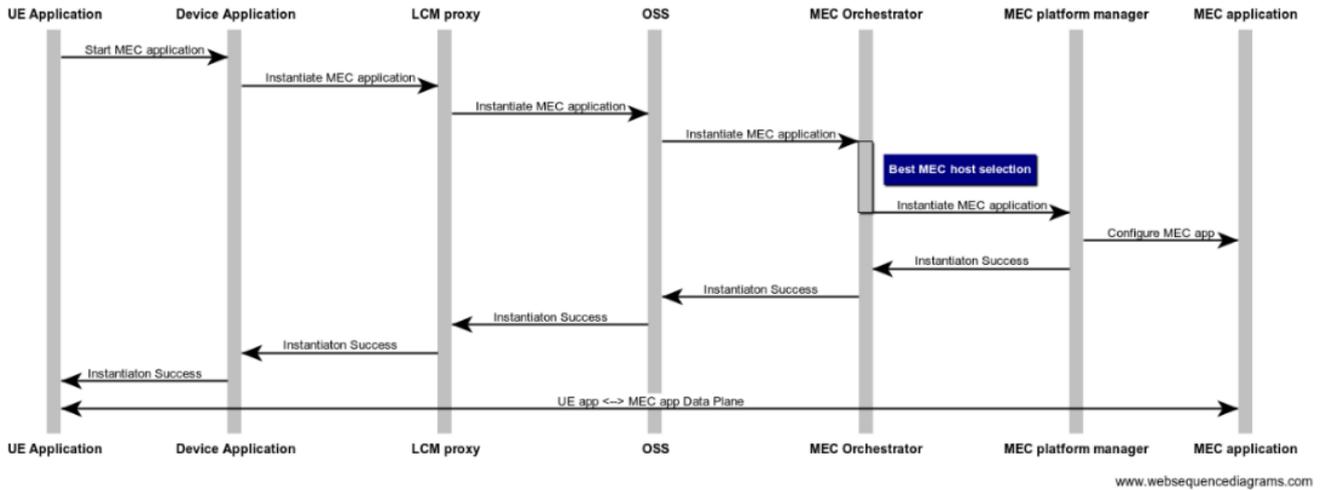

Figure 5 - Sequence diagram of the instantiation of a MEC app on behalf of the user.

The ETSI standard also defines a set of *reference points* [34], i.e. standardized interfaces between the entities of a MEC system. For instance, the Mx2 reference point is used by the device application to request the MEC system to perform operations on a MEC app. The UALCMP is in fact the point of termination of the Mx2 interface towards the device application. The Mp1, between the MEC platform and the MEC app, allows MEC services management (registration and discovery).

## 3. An Architecture for Rapid Prototyping of MEC-based Applications

This section presents our framework for rapid prototyping of MEC apps. Our framework is based on Simu5G, which is enhanced with models of the MEC components described in the previous section. The rationale behind this choice (instead of, say, a standalone model of the MEC system) is twofold: on one hand, Simu5G simulates the 4G/5G data plane, which allows a MEC app developer to obtain a reliable estimate of the performance of her app in a mobile network. On the other hand, Simu5G produces those data that a MEC platform uses to provide MEC services (notably, location and radio information), hence allows a developer to test MEC apps that consume those services. A design choice for our framework is to implement in a standard-compliant way (large subsets of) the MEC reference points towards applications, i.e. the Mx2 and Mp1 interfaces. This design choice allows a developer to interact with our framework in the same way it would with a real MEC system: this way, one can interface production-level code with it. Since Simu5G can run as both a simulator and an emulator, and can interface with external code, a developer has the option of writing parts of her MEC app either as Simu5G modules (e.g., a stub UE app for testing the MEC app) or as standalone external applications. In this section, we describe how we modelled MEC entities. In particular, we focus on our model of the MEC system-level entities, the MEC host, MEC platform and services, and application endpoints.

### 3.1. Model of MEC system-level components

Our framework models both the UALCMP and the MEC orchestrator. The UALCMP is an OMNeT++ module, which includes the TCP/IP stack and the application that implements the Mx2 reference point [36]. With this interface, device applications can query the instantiation and the termination of MEC apps, exposed through a RESTful API. Such requests are then forwarded to the orchestrator which, after handling them, returns a response on the outcome. The UALCMP can also be queried by a device application running outside the simulator, if connected to an INET external interface. Note that – coherently with the ETSI standard – we also allow a device application to join an already instantiated MEC app, in a many-to-one communication model. In this case, the UALCMP simply retrieves and returns the endpoint of such MEC app. For ease of exposition, we do not discuss this possibility further, assuming one-to-one communication henceforth.

The MEC orchestrator is a simple module, connected to the UALCMP. It is configured with a list of the attached MEC hosts. Upon receiving a request for the instantiation of a MEC app, the orchestrator chooses the most suitable MEC host, among those associated to it, according to a user-definable policy which may take into account application requirements (e.g., CPU, memory, disk, required MEC services). Next, it contacts the MEC platform manager of the chosen MEC host in order to eventually trigger the MEC app instantiation, following admission control. A user interested in defining or testing MEC host selection policies can do so by overriding the `chooseBestMECHost()` method of the MEC orchestrator module. The time needed to accomplish the instantiation and termination operations can also be configured.

Interactions between the MEC orchestrator and the UALCMP or the MEC host-level entities are not implemented to be standard-compliant. Instead, they occur using OMNeT++ message passing between modules. This allows us to implement standard-compliant behavior with a simpler, more manageable implementation, without any loss of functionality towards the application endpoints. Moreover, we remark that not all reference points have been standardized at the time of writing, e.g., the interface between the UALCM proxy and the Operation Support System. For this reason, we did not implement the latter, and modelled the above interactions with a configurable delay in the MEC orchestrator.

Finally, note that our framework allows one to instantiate several MEC system, each one with its own MEC orchestrator and set of hosts.

### 3.2. Model of the MEC host-level components

The MEC host is the main building block of the MEC host level architecture. As shown in Figure 6, it includes both management entities, such as the MEC platform Manager and the Virtualisation Infrastructure Manager, and the modules required to run the MEC apps, i.e. the virtualisation infrastructure and the MEC platform. Hereafter, we first describe how the MEC host runs MEC apps. We defer describing the MEC platform and MEC services to the next subsection.

A MEC host comes with a set of hardware resources, namely a processing rate (measured in instructions per second), a given amount of memory and disk space. The main duty of a MEC host is to run MEC apps. As outlined before, a MEC app can be either *external* to our framework, i.e. an external application exchanging information with our framework in real time, or *internal* to our framework, i.e., an OMNeT++ module compiled within it. In the *external* case, the MEC host has an external interface towards the MEC app, which runs somewhere else (e.g., on a virtual machine in a desktop computer connected to our framework via a local network). Accordingly, the speed at which that MEC app runs depends on the hosting hardware (e.g., its CPU speed). In the *internal* case, i.e., when MEC apps run within our framework, instead, we need to manage the pace a MEC app is executed. In a discrete-event simulator, such as Simu5G, simulated time passes between scheduled events. Messages and packets are events, hence a packet exchange (e.g., a request/response pattern) occurs over a span of simulated time. On the other hand, the MEC application logic is embedded within event handlers, and event handling per se does not consume simulated time: any processing done by such applications is thus instantaneous in simulated time. So, to obtain a more realistic model of the processing operations in a MEC app, we modelled the execution time of a block of code by adding an event representing the required CPU computation, which in turn depends on the level of CPU contention at the MEC host. A MEC host can schedule *internal* MEC apps according to two paradigms:

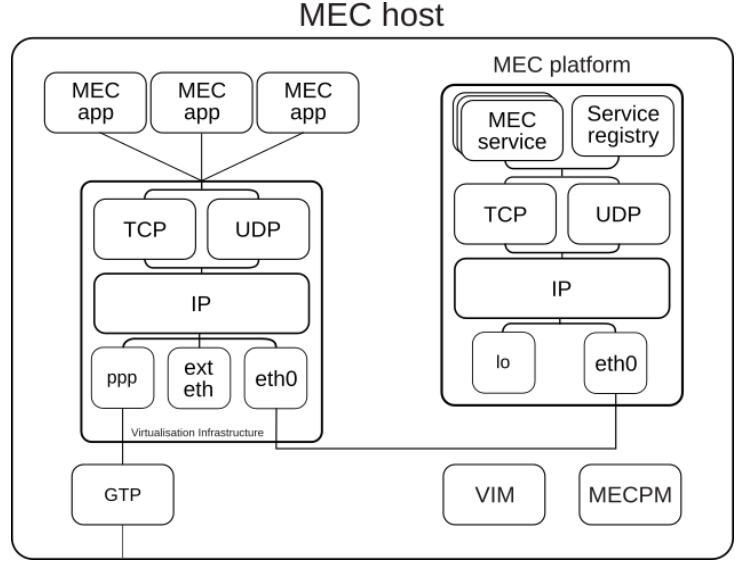

Figure 6 - MEC host modeling within Simu5G

- *Segregation*, whereby each MEC app obtains exactly the amount of computing resources it has stipulated at the time of admission control, even when no other MEC apps are running concurrently;
- *Fair sharing*, whereby active MEC apps share all the available computing resources proportionally to their requested rate, possibly obtaining more capacity than stipulated when contention is low.

At the time of instantiation, a device application requests a *computation rate* $r_i$ for the MEC app. The admission control done by the orchestrator checks (among other things) that $\sum_{j=1}^{n} r_j \leq R$, where $R$ is the MEC host's processing rate. With segregation, a MEC app computation will be served at a rate $r_i$, regardless of the processing contention on the MEC host. This models non-work-conserving scheduling of the MEC app virtual machines on the MEC host. With fair sharing, instead, a MEC app computation will run at a rate

$$r'_i = r_i \cdot \frac{R}{\sum_{j=1}^{n} r_j},$$

where the summation at the denominator includes all the MEC apps that are active when the computation is requested. It is obviously $r'_i \geq r_i$, equality holding only when the admission control limit is reached. Fair sharing approximates Generalized Processor Sharing (GPS) scheduling on the MEC host [30]. The approximation is due to the fact that the actual processing rate $r'_i$ is computed *once* when a computation is requested by the MEC app, rather than updated continuously over time. In fact, $r'_i$ changes over time (because other MEC apps start and terminate, modifying $r'_i$), hence the time at which a computation terminates cannot be calculated once and for all *ex ante*, but must instead be updated every time the summation at the denominator changes. This is a well-known problem with GPS emulation, where a quadratic complexity is required for exact tracking of finishing times. However, exact GPS emulation would only bring a significant accuracy improvement with very long continuous computations done by a MEC app. Note that one can instantiate a *dummy* MEC app on a MEC host, whose only purpose is to eat up resources (notably, processing speed) at the MEC host, so as to simulate a given processing load in an efficient way.

To model the time spent by computation requests in an *internal* MEC app, we allow a user to use a special call `compute(N)`, where $N$ is the number of instructions to be executed. This adds an event in the future, whose firing time is computed based on $N$, *on* the MEC host processing speed, on the scheduling paradigm and – possibly – on the number of active MEC apps at the time of the call. After such delay, the block of C++ code modelling the computation phase of an *internal* MEC app is executed.

### 3.3. Model of the MEC platform

The MEC platform module has its own TCP/IP stack and is connected to the Virtualisation Infrastructure via an Ethernet connection. It contains MEC services and the Service Registry, which are implemented as TCP applications.

MEC Services interface with MEC apps through a RESTful approach. In our framework, they are implemented as HTTP servers and support both the request-response and subscribe-notify communication models. Incoming requests are queued up in FIFO order and served sequentially. A *notification* FIFO queue is also added (different events may generate notifications simultaneously, hence queueing may actually occur). When both are non-empty, the notification queue takes precedence. A MEC service is itself a point of contention, since several MEC apps may request the same service near-simultaneously. So, we need to model delays at a MEC service in a coherent and scalable way. The service time of an HTTP request or a subscription notification event is simulated with a delay computed by a specific method, `calculateRequestServiceTime()` that the user can modify in order to produce such times according to some attributes, such as request type, the number of parameters, a random distribution. By default, the service time is calculated according to a Poisson distribution with a configurable mean. Thus, the response time of an HTTP request depends on the calculated service time and on the number of requests already queued at the server.

A user may want to test scenarios with heavy contention at a MEC service (e.g., due to a very large number of concurrent MEC apps). In a MEC system, in fact, requests can arrive not only from the MEC apps instantiated in the same MEC host where the MEC service runs, but also from MEC apps deployed in other MEC hosts [34]. Simulating many MEC apps increases the computation overhead in Simu5G, and this may constrain the size of the scenario that can be simulated or emulated in real time. To solve this, we provide an implementation which allows one to simulate arbitrary loads at a MEC service at a constant computation cost. We distinguish *foreground* MEC apps, i.e., those that are instantiated on a MEC system, and *background* MEC apps, i.e. those whose load we want to simulate, without paying the overhead of modelling the application logic itself. We model a MEC service as an M/M/1 queueing system, where the service rate is $\mu$ and the arrival rate of foreground/background requests are $\lambda_f$ and $\lambda_b$, respectively, with $\lambda_f \ll \lambda_b$. Moreover, we assume that foreground and background requests are independent. If $\lambda_f + \lambda_b < \mu$, then the system is stable, and the state probabilities seen by an arriving foreground request are the same as the steady-state probabilities seen by a random observer (PASTA property), which are:

$$p_n = \rho^n \cdot (1 - \rho), \tag{1}$$

where $\rho = (\lambda_f + \lambda_b)/\mu$, and $n \geq 0$ is the number of requests in the queue. When a foreground request arrives at the queue, two cases are given: a) no other foreground requests are in the system, or b) there is already one foreground request in the system. In the first case, then one can extract a value for the number of requests already in the system from (1), let it be $n^*$, and schedule the departure of the arriving foreground request $n^* + 1$ service times in the future. This entails extracting a value from an $(n^* + 1)$-stage Erlang distribution with a rate $\mu$.

Assume instead that a foreground request arrives when there is already another in the system. Let $t_0, t_1$ be the arriving times of the former and current foreground requests, respectively. The number of background requests arrived in $[t_0, t_1[$ is a Poisson random variable with a mean $\lambda_b \cdot (t_1 - t_0)$. Once a value is extracted from that distribution, call it $n^*$, then one can follow the same reasoning as above, and extract the inter-departure time between the two foreground requests from an $(n^* + 1)$-stage Erlang distribution with a rate $\mu$.

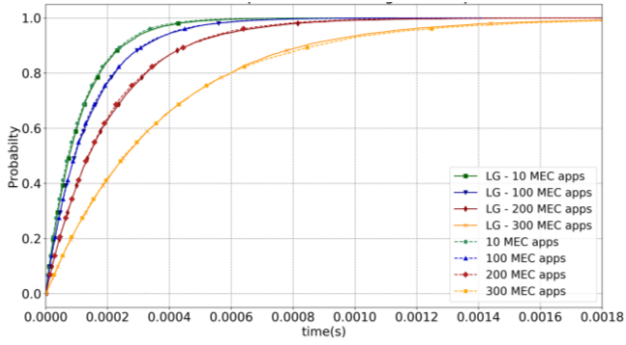 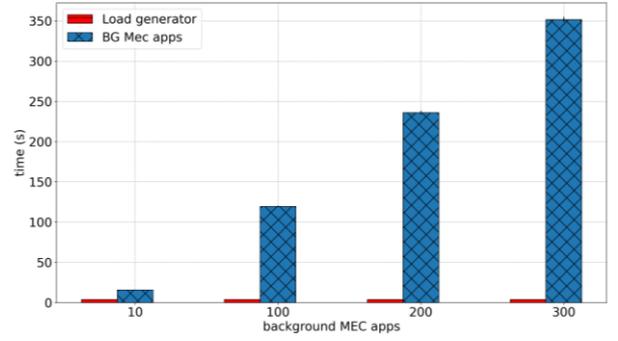

Figure 7 - CDF of the delay of the foreground MEC apps when an increasing number of real/modelled MEC apps interfere at the MEC service.

Figure 8 - Wall-clock time required to simulate 180s of a MEC system, using an increasing number of MEC applications vs. the equivalent load generator

By using the above model, one can store *only* foreground requests in the queue, hence the computation overhead of simulating contention at the MEC service is independent of the background load. If the arrivals and/or services are not exponential, a similar modelling can be used, by substituting the appropriate distributions for the number of requests in the system, the cumulative service time of $n^* + 1$ requests back-to-back, and the number of arrivals within a known interval of time.

To validate the above model, we simulated a simple scenario with a single gNB, a MEC host and three mobile UEs having MEC apps making periodic requests to the Location Service running in the MEC host. The Location Service is loaded by other MEC apps, simulated either as real MEC apps modules or using the above load generator. These send requests exponentially with a rate lambda = 0.024 each, while the three foreground apps send requests every 500ms. The number of background MEC apps varies from 10 to 300 and the simulation time is 180 seconds.

Figure 7 shows the CDF of the response time of *foreground* requests issued by the foreground MEC apps when the number of *background* MEC apps varies. As we can see, results are overlapping – even though requests from foreground MEC apps are periodic, instead of Poisson. However, the computation overhead is quite different. We show in Figure 8 the mean execution time out of 15 independent repetitions, with 95% confidence interval, in the same conditions, on a laptop with an Intel I5 processor. With the load generator, the running time is independent of the number of background MEC apps, and remains fixed at 3.8 seconds. When simulating real MEC apps modules, the running time increases with the number of background MEC apps. Note that, already with 200 real MEC apps, the mean wall-clock time required to run a simulation of 180 seconds exceeds 200 seconds, i.e., simulation time runs *slower than* wall-clock time. This means that emulation would not be workable in that case. By using the above load generator, instead, one could emulate in real time scenarios where MEC services are congested.

## 4. MEC Applications and Services

In this section, we describe how we model the MEC app, device app and UE app application endpoints, showing what configurations are required depending on the possible deployment options (i.e., internal to Simu5G or real applications running outside Simu5G). Moreover, MEC services running within the MEC platform are described, presenting the implementation of two ETSI MEC services.

### 4.1. Model of application endpoints

As already mentioned, Simu5G can also exchange real network packets with external applications. Since our MEC framework implements the Mx2 and Mp1 reference points, a developer can develop and test all of the MEC-related applications as either Simu5G modules, or external applications. In this last case, she can leverage the fact that Simu5G can run in real time to setup live experiments in an emulated environment.

#### 4.1.1. MEC app

Each MEC app must be accompanied by the JSON file describing the Application Descriptor mentioned in Section 2. This file includes the necessary fields to allow on-boarding of an application package into the MEC system, such as:

- *appDId*: unique identifier of the app descriptor;

- *appName*: name of the MEC app;

- *appProvider*: module name necessary for create the OMNeT++ module, if the MEC app is internal the simulator, or it can be left empty if the MEC app is external;

- *appServiceRequired*: the MEC services needed to run the MEC app;

- *virtualComputeDescriptor*: the computation resources required to run the MEC app in a MEC host (i.e. memory, disk and CPU).

If the MEC app is external to Simu5G, a field named *emulatedMecApplication* must be specified. This contains the IP address and port sub-fields identifying the real MEC app endpoint. This way, the MEC orchestrator is made aware that the MEC app to be instantiated runs outside Simu5G, hence it returns the MEC app's endpoint to the device app, instead of instantiating it inside the simulator.

Our framework allows a developer to quickly create a prototype of an internal MEC app, by deriving the *MecAppBase* module. The latter is a base class that manages sockets and OMNeT++ events, leaving to the developer only the implementation of the methods called upon message reception (e.g. from the Service Registry, a MEC service or a UE). For external MEC apps, on the other hand, the only required configuration is the Service Registry's IP address and port, through which the location (i.e. IP and port) of the required MEC services can be discovered.

#### 4.1.2. Device app

To facilitate the task of an application developer, our framework provides a simple device app that can request the installation and termination of a MEC app to the UALCM proxy via the RESTful API implementing the Mx2 reference point. That device app can be contacted by a UE app via UDP socket by means of a simple interface that includes messages for the creation, termination and acknowledgments of a MEC app (e.g. START `mecAppName`, ACK `endpoint`). This interface can be used by internal and external UE apps, thus a developer has only to write the UE app logic and simply query the above device app when a MEC app instantiation (or termination) is needed.

Note that our UALCM proxy also accepts requests from external device apps, since the interface between the two occurs via the ETSI compliant Mx2 API. In this case, the external device app only needs to be configured with the address of the UALCM proxy.

#### 4.1.3. UE app

As far as the UE app is concerned, the same approach as for the other apps has been used, i.e. it can be both internal and external to Simu5G. The first option can be particularly useful when the UE app is meant to run as a stub, e.g. to just issue requests at a predefined rate and record statistics.

```
// Device app endpoint
deviceAppEndpoint = (devAppIP, devAppPort)
UDPSocket devAppSocket;
// request MEC app instantiation
devAppSock.sendTo(deviceAppEndPoint,"START MecAppName")
mecAppEndpoint = devAppSock.recv()
// {
//    UE app logic
// }
// request MEC app termination
devAppSock.sendTo(deviceAppEndPoint,"STOP MecAppName"
```

Figure 9 - Pseudo-code to execute a UE app in a MEC system

Figure 9 describes the UE app pseudocode necessary to request the instantiation (and termination) of the MEC app, via the built-in device app available in Simu5G. The above design makes it very easy for a developer to port to a MEC environment a preexisting client-server application: the existing server application can be deployed via the MEC framework as a MEC app, whereas the entire client logic is inserted virtually as is within the curly brackets in Figure 9 with minimal to null modifications.

### 4.2. Model of MEC services

As far as MEC services are concerned, we provide both a general-purpose module for rapidly prototyping ETSI-compliant MEC services, and two useful standard services, namely RNIS and Location Service. Our framework comes with a basic module, called `MecServiceBase`, which implements all the non-functional requirements needed for running an HTTP server, leaving to a developer only to implement the methods to handle HTTP requests and subscriptions of the RESTful API. The interface also maintains the set of eNB/gNB modules associated to the MEC host.

Two standardized MEC services are currently implemented, namely the RNIS and the Location Service. The RNIS is used to gather up-to-date information regarding radio network condition and the UEs connected to the base station associated to the MEC hosts [31]. Such information allows MEC apps to have real-time information on the network performance, which can be used e.g., to provide an improved Quality of Experience (QoE) to end-users: for instance, to decrease the video quality of a streaming application while the channel is poor or the cell is overloaded, thus avoiding video rebuffering. A subset of the RNIS API is implemented, which includes the *Layer-2 measures* resource, reporting gNB Layer 2 measurements, such as packet delay, throughput, number of active UEs with downlink/uplink traffic, data volumes, cell utilization or packet data rate. By varying the network configuration, one can carry out evaluations and validations of MEC apps using the RNIS in different network conditions. The RNIS gets its information from gNB modules within Simu5G. Hereafter we show how this is achieved without modifying the existing models within Simu5G. A dedicated module, named `gNodeBStatsCollector` (collector, from now on), can be added to a gNB to retrieve measures from the NIC modules. This should be instantiated only when necessary (i.e., when a MEC host exposes the RNIS), as it is costly from a processing overhead standpoint. The large overhead is due to the fact that a collector manages several timers used to trigger information-retrieving procedures, and their firing events slow down the execution time. Each logged measure is stored in a `L2Measure` object by the collector. Different aggregators can be used to compute the value to be returned when the RNIS requires it, e.g. average, moving average, or last sample. A user can also implement its own aggregator or configure window timers according to documents [37]-[38]. Some L2 measures of the RNIS involve different sublayers of the NR stack. For instance, some delays are measured from the arrival to the PDCP layer to the reception of the HARQ ACK at the MAC layer. This is handled without modifications to the layers, by adding a new module in the NIC, called `packetFlowManager`, which receives information by the relevant layers when the above events are triggered and maintains the data structures necessary to identify the same payload at different NR sublayers.

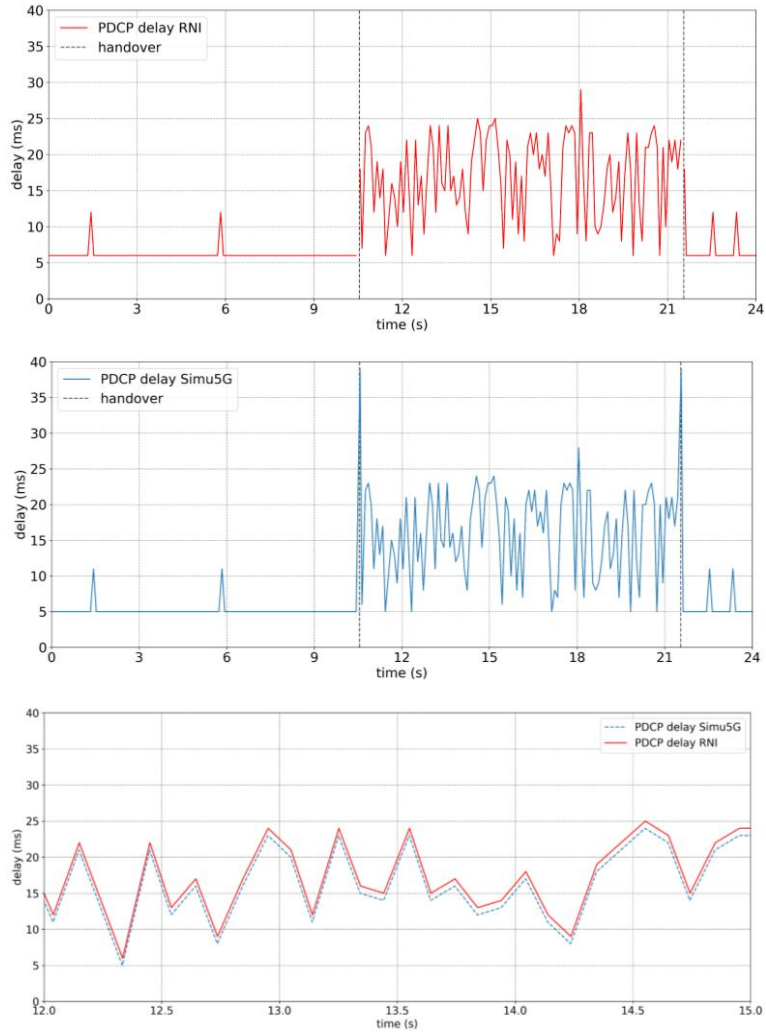

Figure 10 - Downlink packet delay obtained by querying the RNIS (top) or using Simu5G metrics (center), and zoom-in of the difference between the two (bottom) between 12 and 15 seconds.

We validate the results returned by our implementation of the RNIS by comparing them against similar metrics already provided by Simu5G. Figure 10, top, shows the PDCP packet delay in downlink, returned by the `packetFlowManager`, of a UE that connects to three gNBs over time, the middle one having other 100 UEs periodically receiving packets from a server. In order to show readable and interpretable results, the number of available 5G resource blocks has been downsized to 10. When the UE handovers to the middle gNB (handovers are marked by dashed vertical lines), the delay increases. Figure 10, center, depicts instead the PDCP delay calculated by Simu5G in the same simulation. It can be observed that the patterns are correlated, suggesting a correct implementation of the RNIS metric. A closer look, shown in Figure 10, bottom, and reporting a zoom of the two above graphs in a single reference, shows that there is a constant difference of 1ms, which is due to the fact that the RNIS uses HARQ ACK reception to mark the end of the measurement interval, and that occurs 1ms after the UE has received the packet over the air, the latter being the event when the Simu5G statistic is computed instead. Moreover, delay at handover times are different (as per the spikes in Figure 10, center, around $t = 10$, $t = 21$), since Simu5G starts measuring the delay at the entry *within the NR*

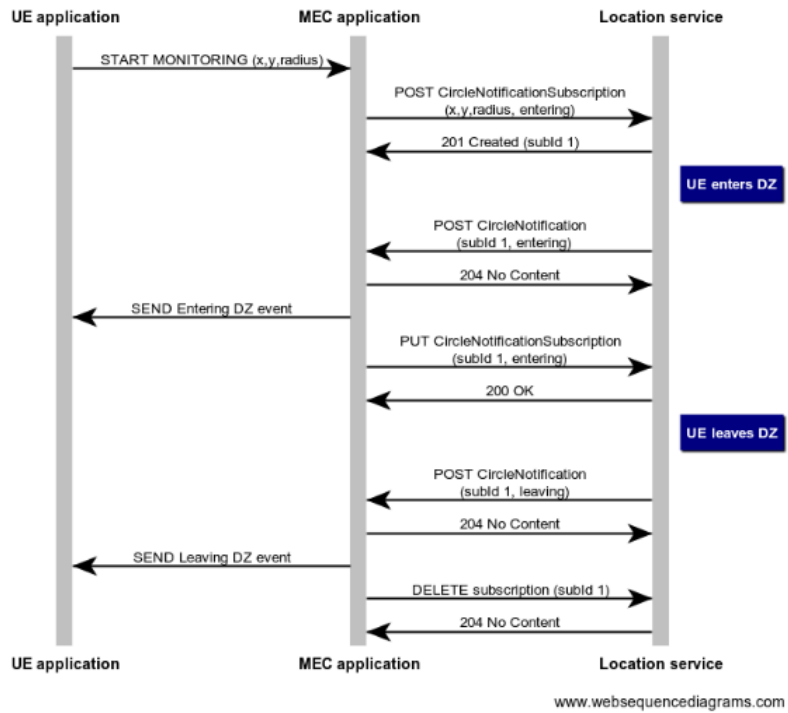

Figure 11 - Sequence diagram of the danger zone Warning Alert use

*stack*, whereas the RNIS starts at the entry *at the PDCP layer*. The two are usually the same, except during handover, when Simu5G adds the handover delay to the computation.

The Location Service provides accurate information about UE and/or base station position, enabling active device location tracking and location-based service recommendation. The reference API is described in [32] and it is based on the RESTful API originally defined by the Open Mobile Alliance [39]. UE positions are stored and periodically updated on the gNBs to which they are currently connected and are expressed as three-dimensional Euclidian coordinates provided by the INET Mobility model library. A MEC app can query the position of the UEs with different granularities: a single UE, a group of UEs, only the UEs connected to a specific base station or a group of them. The "UE Area subscription" is also present [32]. It follows the "Area (circle) location notification" subscription defined in [39], on which a MEC app subscribes to receive notifications when a UE *enters* or *leaves* a circular zone described by its center coordinates and radius.

## 5. A Case Study

We describe here an example of a distributed application, where both endpoints (UE app and MEC app) are external. In our scenario, a vehicular UE moves in a simulated floorplan, and wants to be notified when it enters a *danger zone*, i.e. a black-ice area. To do so, it sets up a MEC app that uses the Location Service to check the UE position. The network scenario is composed of a gNB with some associated UEs, i.e. cars equipped with a NR interface, moving towards the danger zone. A MEC host is attached to the gNB and runs the Location Service. The latter provides the "Area (circle) notification subscription" resource [32]. On that resource, a client requests to monitor when a UE enters in a circular zone.

More specifically, after the MEC app instantiation, the UE app requests it to monitor a specific zone. We assume the UE starts outside the danger zone, hence the first subscription is for the *entering*-danger zone event. When the MEC app receives the *entering* notification, it modifies its subscription, to be notified when it *leaves* the danger zone. The UE app is only informed upon notification events. The resulting sequence diagram is shown in Figure 11.

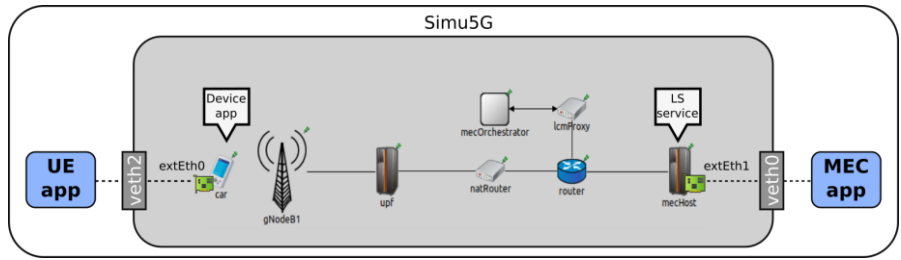

Figure 12 - Simu5G set up for emulation

```
# create virtual ethernet link: veth0 <--> veth1, veth2 <--> veth3
sudo ip link add veth0 type veth peer name veth1
sudo ip link add veth2 type veth peer name veth3

# veth0 <--> veth1 link uses 192.168.3.x addresses
sudo ip addr add 192.168.3.2 dev veth1
# veth2 <--> veth3 link uses 192.168.2.x addresses
sudo ip addr add 192.168.2.2 dev veth3

# bring up both interfaces
sudo ip link set veth0 up; sudo ip link set veth1 up;
sudo ip link set veth2 up; sudo ip link set veth3 up

# add routes for new link
sudo route add -net 192.168.3.0 netmask 255.255.255.0 dev veth1
sudo route add -net 192.168.2.0 netmask 255.255.255.0 dev veth3

# UE's eth interface is 192.168.4.1, route for Device app
sudo route add -net 192.168.4.0 netmask 255.255.255.0 dev veth1
# natRouter left interface (to UE) address is 10.0.2.1
sudo route add -net 10.0.2.0 netmask 255.255.255.0 dev veth1
# natRouter right (to MEC host) interface address is 10.0.3.2
sudo route add -net 10.0.3.0 netmask 255.255.255.0 dev veth3
# mecPlatform interface address is 10.0.5.2
# (for MEC services and Servive registry)
sudo route add -net 10.0.5.0 netmask 255.255.255.0 dev veth3

# run simulation
simu5g -u Cmdenv -c ExtClientServer_Socket
```

Figure 13 - *veth* interfaces configuration

We run the above scenario in an emulated environment on which both the UE app and the MEC app run outside the simulator, on the same host where Simu5G runs, as shown in Figure 12. The device app, instead, resides on the UE module inside Simu5G. We now describe the configurations needed to execute this testbed on a computer running Linux Ubuntu operating system.

First of all, the network scenario must be configured within Simu5G. For the low-level details, we refer the interested reader to papers describing Simu5G configuration (e.g., [6] and [1] – describing configuration of MEC scenarios) as well as the website documentation [7]. Here, we limit ourselves to recalling that Simu5G modules can exchange packets with outside applications using *ExtLowerEthernetInterface* OMNeT++ modules. These must be inserted in the car and *mecHost* modules, so as to inject/receive UE app and MEC app traffic inside the simulator. In our case, these interfaces are created as Virtual Ethernets (*veth*).

Then, the host operating system must be instructed to route packets into Simu5G via the *veth* interfaces by adding ad-hoc routing rules to reach the device app and the MEC platform, and to enable the transport of the traffic between the UE app and the MEC app through the emulated 5G network. Since the MEC app and the UE app run on the same host, a mechanism to bypass the operating system and steer the traffic towards the simulator is needed. A possible solution is to insert a *natRouter* module in the Simu5G network. This way, both real applications send packet to the IP addresses of the interfaces of the *natRouter*, which in turn performs Network Address Translation by changing the destination addresses to the proper real application's addresses. Figure 13 shows the commands required to create, configure and add routes of the *veth* interfaces on a host equipped with Linux Ubuntu 18.04 operating system. Moreover, routing information within the simulator have to be added, as the OMNeT++ platform does not know autonomously how to route packets to destinations outside the simulated/emulated network.

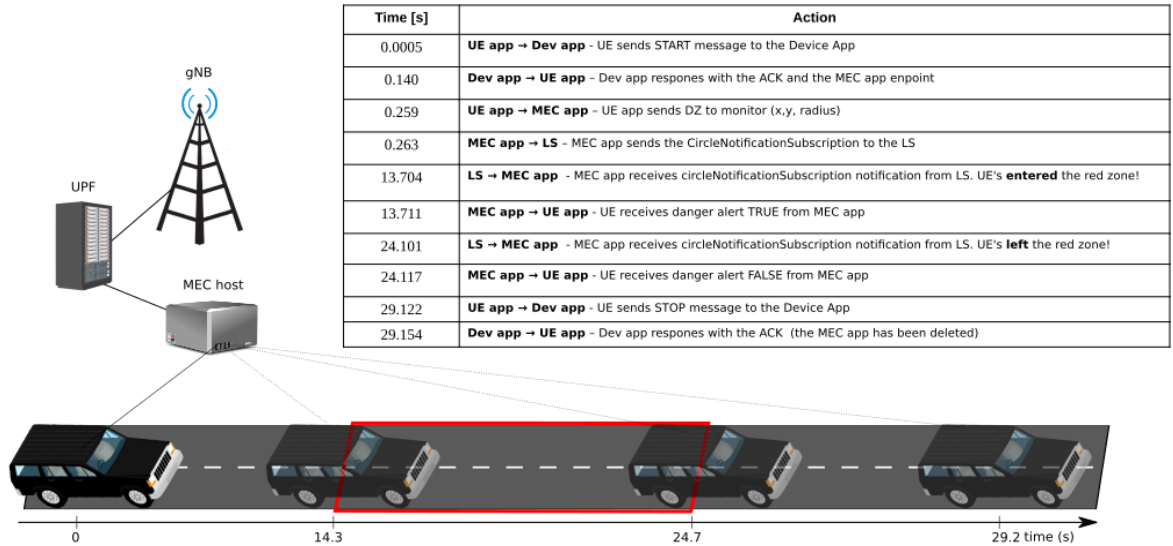

Figure 14 - Timeline of the message events during the execution of the use case

After the networking has been set, the last part to be configured is related to the MEC app in use, which is the *MECWarningAlertApp* in this case. An *appDescriptor* file describes the MEC app in terms of its configuration parameters. For example, for the purpose of the emulation, the *emulatedMecApplication* field informs the MEC orchestrator about the IP address and port where the MEC app listens for UE app requests. Once the UE app and the MEC app know the device app and the Service Registry endpoints respectively, the testbed is ready to run.

Finally, Figure 14 shows the timeline of the car moving towards the danger zone, with the relative events captured by the Wireshark tool monitoring the *veth0* and *veth2* interfaces.

As can be seen, the configuration effort is modest, and mostly is focused on the networking settings. This is necessary for any scenario where OMNeT++ is run in emulated mode and does not depend on the MEC framework. The latter only requires the creation of the descriptor file for the MEC app and the creation of the network scenario to be executed.

## 6. Related Works

In this section we review the available works on 5G simulation, MEC simulation and MEC services testing.

As far as 5G simulators are concerned, there are both those that simulate the physical layer, such as [14]-[15], and those that allow end-to-end, application-level simulations, like Simu5G. In this last category we find 5G-Lena [16] and 5G-air-simulator [17]. As far as we know, neither supports MEC or allow real-time emulation of a 5G network.

A critical review of the Fog/Edge simulators and emulators available to date is reported in [24]. The tools described in the paper have been developed to evaluate computing infrastructures in terms of deployment scenarios, energy consumption and operational costs. They address Internet of Things application and resource management in an Edge/Fog/Cloud continuum. The most widely used seems to be iFogSim, although it does not take network aspects into account too much. The underlying network condition, with increasing QoS requirements from users, has become of paramount importance and therefore needs to be carefully considered as well. More importantly, these tools do not address ETSI MEC and the related functionalities, such as MEC services, that can retrieve information about the underlying network for the edge applications, i.e. MEC apps.

The work most closely related to ours seems to be the MEC simulator described in [23]. Devised for the ns3 framework, it allows one to simulate simplified models of distributed MEC-based applications over a MEC system, using non ETSI-compliant interfaces. The simulator includes models of MEC orchestrator and app mobility. There are, however, several crucial differences with our work: first, one cannot use real MEC-based applications and run them on this framework, as it does on Simu5G. Second, there seems to be no models for MEC services and MEC platform: the simulator, in fact, includes *dummy* base stations, whose only role is to have UEs associated to them on a proximity basis, without modelling radio communication (in fact, RAN latency in the experiments described in [23] is modelled as a probability distribution). Therefore, on one hand, these base stations cannot provide any radio information to a MEC platform (e.g., the level of resource occupation, or the channel quality of a mobile); on the other hand, such an architecture also prevents straightforward interoperability with the 4G/5G libraries already available for ns3 (such as 5G-Lena), where the above information could be generated. Last, but not least, to the best of our knowledge, [23] cannot run real-time emulations.

All the above works focus on MEC modelling, and either neglect or abstract away the role of the underlying RAN. We remark that the interplay between the RAN and the MEC does not limit to the former transporting packets generated or consumed by the latter. Rather, and no less importantly, the RAN generates the information that MEC apps use via MEC services (notably, those returned by the RNIS). While it could make at least some sense to abstract away the modelling of RAN communication impairments (e.g., just a delay distribution and a bandwidth), it is just pointless to have the RNIS without an underlying credible and *detailed* model of the RAN. We also remark that an architecture including both the 5G access and the MEC system allows one to test the impact of MEC traffic on the network.

The ETSI group also provides a portal for exploring the functionalities of MEC services. It is called "ETSI MEC Sandbox" [41] and offers an environment on which the users can choose among different real-time access network scenarios (e.g. 4G, 5G, WiFi) and practice with ETSI MEC service APIs. RESTful resources can be queried from both browser and existing applications testing MEC application use cases. However, the available settings are limited and the network's parameters, like fading, cell interferences and multi-carrier components cannot be modified in order to produce different behaviors, limiting the use of this tool to the sole purpose of experimenting with the MEC services API, without allowing one to customize the network scenarios.

The OpenAirInterface 5G-RAN project [22] is under way as we write, with a schedule foreseen to complete in the second half of 2022. The project promises to deliver an open-source software implementation of a programmable 5G RAN. If and when this project is completed, it will probably be a good tool to experiment with the interplay between MEC and the RAN: for instance, one may envisage building a system where OAI 5G-RAN interacts with a MEC infrastructure (e.g., Intel OpenNESS [28]), and UEs can enjoy MEC services in an emulated environment. However, this requires OpenAirInterface 5G-RAN to interface with the MEC system to export the information needed for MEC services, and we have no information that this is included in the current plan of the OpenAirInterface consortium. Even if it was, we are somewhat skeptical that such an environment would provide a MEC app developer with the possibility to test its own MEC apps quickly and under controlled conditions, possibly including large-scale communication or congestion at the MEC host.

Finally, three previous works of ours [13],[2],[27] are related to this one. Work [13] describes the MEC model developed for SimuLTE. This work is meant to be used for research purposes, i.e. to allow a SimuLTE user to instantiate simplified models of MEC-based applications – similarly to [23], and evaluate their impact on the data plane. While the work described in this paper reuses code from [13], suitably adapted to the 5G environment, its purpose is to serve MEC developers, rather than networking researchers. Accordingly, it has been re-engineered and enhanced with models of ETSI-MEC components, such as the UALCMP or the device app, that are necessary for seamless emulation of MEC functionalities for a developer, and standard-compliant external interfaces to the application world have been implemented. For instance, the use case described in Section 5 could not have been emulated in that environment – not even substituting 4G access for 5G. Our paper [2] describes the real-time emulation capabilities of an early release of Simu5G. As a case study, we show therein that Simu5G can provide 5G transport to a MEC app running on Intel OpenNESS [28]. On one hand, the current version of Simu5G is quite different from the one described in [2], especially for what concerns real-time emulation (thanks to the new versions of OMNeT++ and INET). On the other hand, the MEC app described therein is a simple client/server video application running on a virtual machine instantiated on a MEC host. All the management-plane interactions are missing, and the service is statically configured offline. In [27], we used Simu5G together with and Intel® CoFluent™ studio [26] to evaluate the performance of MEC apps running over different 4G/5G deployments. CoFluent is a modelling and simulation tool for optimizing, analyzing and predicting the performance of complex systems, that models and simulates HW/SW systems with microinstruction-level accuracy. In [27], the MEC host was modelled within CoFluent, and a video-streaming MEC app was modelled within it. However, the co-simulation framework described in [27] is hardly comparable to what we describe in this paper. In fact, it is based on file exchanges between Simu5G and CoFluent, which run separately. This allows one to compute the round-trip delay, including both communication and computation. However, no feedback between the two simulators can exist (e.g., network conditions influencing the MEC app behavior, e.g. via RNIS, or user behavior depending on MEC app results). This could only be enabled by scheduling *both* CoFluent and Simu5G events in the same unified framework, which would require a considerable amount of work. Moreover, while CoFluent can model a MEC host with high accuracy, its very accuracy makes it unsuitable to model a *large-scale* MEC system, which can instead be modelled with Simu5G.

## 7. Conclusions and Future Work

This work described a framework for rapid prototyping of MEC-based applications. Our framework, based on the Simu5G discrete-event simulator, gives a developer two options: the first one is to write application logic (UE app and MEC app) as Simu5G modules. This is quite simple, and allows the developer to test the application logic in a pre-production stage, obtaining reliable performance metrics in a customizable 5G scenario. The other option is to use existing MEC-based application endpoints, and run them through our framework, which provides not only 5G packet transport, but also MEC signaling functionalities and MEC services. This can also be done in real time, e.g. for demonstration purposes or to test interactions with a human end user or other external software. We have described the modelling of the MEC components in our framework, validated our implementation of MEC services, and showed that one can set up an emulation testbed with external application quite simply, on off-the-shelf hardware. We believe that the work described in this paper will be useful to MEC app developers. To the best of our knowledge, there are no tools with similar functionalities available to the developer community.

At the time of writing, the above framework is being used in the framework of the Hexa-X EU project [29]. In that framework, it will support the development, validation, evaluation and demonstration of federated learning of explainable AI models. More specifically, we plan to use our framework to evaluate network protocols for federated learning, where learning logic can run on both UEs and in MEC systems.

**Acknowledgments**

This work was partially supported by the Italian Ministry of Education and Research (MIUR) in the framework of the Cross-Lab project (Departments of Excellence), and by the European Commission through the H2020 projects Hexa-X (Grant Agreement no. 101015956).